\documentclass{article}
\usepackage{fortschritte}
\usepackage{epsfig,multicol,multirow}
\usepackage{amsmath,subfigure,latexsym,amssymb}
\usepackage{fps}
\def\beq{\begin{equation}}                     % 
\def\eeq{\end{equation}}                       %
\def\bea{\begin{eqnarray}}                     %         %
\def\eea{\end{eqnarray}}                       %       % 
\def\lsi{\raise0.3ex\hbox{$<$\kern-0.75em\raise-1.1ex\hbox{$\sim$}}}
\def\gsi{\raise0.3ex\hbox{$>$\kern-0.75em\raise-1.1ex\hbox{$\sim$}}}

\def\backder{\raise1.4ex\hbox{$\leftarrow$\kern-0.75em\raise-1.4ex\hbox{$\partial$}}}

\newcommand{\lsim}{\mathop{\lsi}}

\newcommand{\backderi}{\mathop{\backder}}
\newcommand{\nn}{\nonumber}
\newcommand{\NN}{{\kern+.25em\sf{N}\kern-.78em\sf{I} \kern+.78em\kern-.25em}}

\begin {document}                 
\def\titleline{
%%%%%%%%%%%%%%%%%%%%%%%%%%%%%%%%%%%%%%%%%%%%%%
%                                            
% Insert now the text of your title.         
% Make a linebreak in the title with         
%                                            
%            \newtitleline                   
%                                            
%%%%%%%%%%%%%%%%%%%%%%%%%%%%%%%%%%%%%%%%%%%%%%%
Numerical Results for $U(1)$ Gauge Theory on \\        %         %
2d and 4d Non-Commutative Spaces              %       %
%                                             %     %%%%%%%%%%%%%%
%                                             %       %
%%%%%%%%%%%%%%%%%%%%%%%%%%%%%%%%%%%%%%%%%%%%%%%         %
}
\def\email_speaker{
{\tt 
%%%%%%%%%%%%%%%%%%%%%%%%%%%%%%%%%%%%%%%%%%%%%%
%                                                  
% Insert now the e-mail address of the speaker or  
% the author that should get the electronic mail   
% of the publishing house                           
%                                                  
%%%%%%%%%%%%%%%%%%%%%%%%%%%%%%%%%%%%%%%%%%%%%%         %
bietenho@physik.hu-berlin.de                 %       %
%                                            %     %%%%%%%%%%%%%%
%                                            %       %       
%%%%%%%%%%%%%%%%%%%%%%%%%%%%%%%%%%%%%%%%%%%%%%         %
}}
\def\authors{
%                                            
%  Insert now the name (names) of the author 
%  (authors).                                
%  In the case of several authors with       
%  different addresses use labels e.g.       
%                                            
%             \1ad  , \2ad  etc.             
%                                            
%  to indicate that a given author has the   
%  address number 1 , 2 , etc.               
%  Signify the person with the above e-mail  
%  address with \sp (behind the address      
%  labels, e.g.                              
%                                            
%            \2ad\sp                         
%                                            
%  when the 2nd author                       
%  happens to be the 'speaker')                                  
%%%%%%%%%%%%%%%%%%%%%%%%%%%%%%%%%%%%%%%%%%%%%%         %
W. Bietenholz \1ad, A. Bigarini \2ad, F. Hofheinz \1ad, \\ 
J. Nishimura \3ad, Y. Susaki \3ad \4ad and J. Volkholz \1ad\sp 
%%%%%%%%%%%%%%%%%%%%%%%%%%%%%%%%%%%%%%%%%%%%%%         %
}
\def\addresses{
%%%%%%%%%%%%%%%%%%%%%%%%%%%%%%%%%%%%%%%%%%%%%%
%                                            
% List now the address. In the case of       
% several addresses list them by numbers     
% using e.g.                                 
%                                            
%             \1ad , \2ad   etc.             
%                                            
% to numerate address 1 , 2 , etc.           
%                                            
%%%%%%%%%%%%%%%%%%%%%%%%%%%%%%%%%%%%%%%%%%%%%
\1ad                                        %          
Insistut f\"{u}r Physik, Humboldt Universit\"{a}t zu Berlin \\
Newtonstr.\ 15, D-12489 Berlin, Germany \\
\2ad Dipartimento di Fisica, Universita' degli Studi di Perugia,\\
and INFN, Sezione di Perugia, Via Pascoli 1, I-06100 Italy\\ 
\3ad High Energy Accelerator Research Organization (KEK)\\ 
1-1 Oho, Tsukuba 305-0801, Japan \\
\4ad Institute of Physics, University of Tsukuba, \\
Tsukuba, Ibaraki 305-8571, Japan \\
%%%%%%%%%%%%%%%%%%%%%%%%%%%%%%%%%%%%%%%%%%%%%
}
\def\abstracttext{
%%%%%%%%%%%%%%%%%%%%%%%%%%%%%%%%%%%%%%%%%%%%%
%                                             
% Insert now the text of your abstract.       
%                                             
%%%%%%%%%%%%%%%%%%%%%%%%%%%%%%%%%%%%%%%%%%%%%         %
We present non-perturbative results for $U(1)$ gauge theory
in spaces, which include a non-commutative plane. In contrast to
the commutative space, such gauge theories involve a Yang-Mills term,
and the Wilson loop is complex on the non-perturbative level.
We first consider the 2d case: small Wilson loops follows an area law,
whereas for large Wilson loops the complex phase rises linearly with the area.
In four dimensions the behavior is qualitatively similar for loops
in the non-commutative plane, whereas the loops in other planes follow
closely the commutative pattern.
In $d=2$ our results can be extrapolated safely to the continuum limit,
and in $d=4$ we report on recent progress towards this goal.
%%%%%%%%%%%%%%%%%%%%%%%%%%%%%%%%%%%%%%%%%%%%%         %
}
\large
\makefront
%%%%%%%%%%%%%%%%%%%%%%%%%%%%%%%%%%%%%%%%%%%%%%%%
%                                              %
%  Insert now the remaining parts of           %
%  your article.                               %
%                                              %
%%%%%%%%%%%%%%%%%%%%%%%%%%%%%%%%%%%%%%%%%%%%%%%%
\vspace*{-10mm}
\section{Non-commutative $U(1)$ gauge theory}

We consider the simplest version of non-commutative (NC) spaces:
only two (Euclidean) coordinates are NC, and the non-commutativity
parameter $\theta$ is constant,
\beq
[ \hat x_{i}, \hat x_{j} ] = i \, \theta \, \epsilon_{ij} \qquad
({\rm indices}~~ i,j \in \{ 1,2 \}) \ .
\eeq
According to a historic observation by Peierls \cite{Pei33}, 
such coordinates describe a charged particle moving in a 
(commutative) plane,
which is crossed by a strong, orthogonal magnetic field $B$. 
In fact, if we ignore the kinetic term in the Lagrangian, such a 
particle has the canonical momentum $p_{i} = q B \epsilon_{ij} x_{j}$,
where $q$ is the electric charge. 
Under canonical quantization we obtain
\beq
[\hat x_{i} , \hat p_{i} ] = i \hbar = 
q B \epsilon_{ij} [ \hat x_{i}, \hat x_{j} ] \ ,
\eeq
which is a NC relation with $\theta \propto 1/B$.
Based on this idea, NC field theory plays a r\^{o}le as a formalism
in solid state physics.

A similar idea is also used to map string theory in a magnetic
background onto NC field theory. The construction of this mapping 
\cite{SeiWit} has tremendously boosted the interest in
NC field theories (i.e.\ field theories on NC spaces).

%For a recent review, see for example \cite{SzaboRev}.
An important mathematical result states that we can return to the use of
ordinary (commutative) coordinates if all the fields are
multiplied by {\em star products},
\beq
\phi (x) \star \psi (x) := \phi (x) \exp \Big( \, 
\frac{1}{2} \backderi \,\! _{i} \, \theta \, \epsilon_{ij} \,
\vec \partial_{j} \, \Big) \ \psi (x) \ .
\eeq
Here we focus on pure $U(1)$ gauge theory, which has the 
Euclidean action
\bea
S[A] &=& \frac{1}{4} \int d^{4}x \, F_{\mu \nu} \star F_{\mu \nu} \ , \nn \\
F_{\mu\nu} &=& \partial_{\mu} A_{\nu} - \partial_{\nu} A_{\mu}
+ ig [ A_{\mu},A_{\nu}]_{\star} \ ,
\eea
where the last term is a star-commutator. This action is star-gauge
invariant, i.e.\ invariant under transformations
\beq
A_{\mu}(x) \to U(x) \star A_{\mu}(x) \star U(x)^{\dagger}
- \frac{i}{g} U(x) \star \partial_{\mu}U(x)^{\dagger} \ ,
\eeq
where $U(x)$ is star-unitary, $U(x)^{\dagger} \star U(x) = 1 \!\! 1$.

Other $U(N)$ gauge theories 
may be studied along the same lines, but the formulation of $SU(N)$
gauge theories runs into trouble on NC spaces. To see this problem,
let us represent the gauge field as $A_{\mu}^{a} T_{a}$,
$T_{a}$ being the Hermitian generators of the gauge group.
Then the star commutator can be decomposed as
\beq
2 \ [ A_{\mu}^{a} T_{a} ,  A_{\nu}^{b} T_{b} ]_{\star} =
[ A_{\mu}^{a} , A_{\nu}^{b} ]_{\star} \cdot
\{ T_{a},  T_{b} \}
 + \{ A_{\mu}^{a} , A_{\nu}^{b} \}_{\star} \cdot
[ T_{a}, T_{b}] \ .
\eeq
If we deal with $SU(N)$ gauge theory we have the additional condition
${\rm Tr} \, T_{a} =0$, which should be reproduced by the 
star-commutator for the algebra to close. In the commutative case,
$\theta = 0$, the first term vanishes 
( $[ A_{\mu}^{a} , A_{\nu}^{b} ]=0$ ). Then the algebra closes because
${\rm Tr} \, [T_{a}, T_{b}]$ trivially vanishes.
For finite $\theta$, however, also the first term contributes, which
causes trouble (because ${\rm Tr} \, \{ T_{a}, T_{b} \} \propto \delta_{ab}$ 
does not vanish).

Therefore it is motivated to concentrate on $U(1)$ as a physical gauge group,
which can be accommodated on a NC space.

\section{Mapping onto a twisted Eguchi-Kawai model}

As a first step towards a formulation to be used in Monte Carlo
simulations, we introduce a lattice structure in the NC plane. Since all
space-points are somewhat fuzzy, we cannot expect sharp lattice sites, 
but the operator identity
\beq
\exp \Big( i \frac{2 \pi}{a} \hat x_{i} \Big) = \hat 1 \!\! 1 
\eeq
does nevertheless impose a lattice structure with lattice spacing $a$.
If we require the momentum components $k_{i}$ to be periodic over 
the Brillouin zone, the above condition implies that only discrete 
momenta occur (see e.g.\ \cite{SzaboRev}), 
which is characteristic for a finite volume.
If we assume a periodic $N \times N$ lattice, the allowed
momenta are separated by \
$\Delta k_{i} = \frac{2\pi}{aN}$, \ and the
non-commutativity parameter can be identified as
\beq
\theta = \frac{1}{\pi}N a^{2} \ .
\eeq
Therefore we are interested in a {\em double scaling limit}
$ a \to 0$, $N \to \infty$ with $Na^{2} = const.$, which leads to
a continuous non-commutative plane of infinite extent.

However, even on the lattice it is far from obvious how to simulate NC 
gauge theory; note that this seems to require star-unitary link
variables. It is highly profitable to map the system (or its NC part)
onto a {\em twisted Eguchi-Kawai model}. The latter is defined on a single 
space-point and its action has the form \cite{TEK}
\beq
S_{\rm TEK} [U] = - N \beta \sum_{i \neq j} Z_{ij} {\rm Tr} \,
\Big( U_{i} U_{j} U_{i}^{\dagger} U_{j}^{\dagger} \Big) \ .
\eeq
$U_{1}$ and  $U_{2}$ are unitary $N \times N$ matrices which encode the degrees
of freedom of the lattice model, and the twist factor is chosen here
as $Z_{21} = Z_{12}^{*} = \exp ( \pi i (N+1)/N)$, where $N$ has to be odd.
Then there is an exact equivalence to the lattice NC gauge theory,
i.e.\ the algebras are fully identical, as Ref.\ \cite{AIIKKT} showed
in the large $N$ limit. A refined consideration found such a mapping
even at finite $N$ \cite{AMNS}. This is the form which is suitable
for numerical simulations.

It is straightforward to formulate (the analogue of)
Wilson loops in the framework of the matrix model,
\beq
W_{ij} ( I \times J) = \frac{1}{N} Z_{ij}^{\pm IJ} \
{\rm Tr} \, \Big( U_{i}^{I} U_{j}^{J} U_{i}^{\dagger \, I} 
U_{j}^{\dagger \, J} \Big) \ .
\eeq
This corresponds to a rectangular Wilson loop of sides $aI$ and $aJ$.
Mapping this quantity back to the lattice leads in fact to
a sensible definition of a Wilson loop in the NC gauge theory \cite{IIKK}.
The exponent of the twist factor is taken positive (negative) for an 
anti-clockwise (clockwise) orientation. The Wilson
loop is complex in general, although this property cannot be seen
perturbatively in $d=2$ \cite{ADM}. For further perturbative studies we refer
to Refs.\ \cite{pertu}, and to Refs.\ \cite{semicla} for semi-classical
approaches. 

%In the planar limit, $N \to \infty$ at fixed $\beta$, this model
%coincides with $U( N \to \infty )$ lattice gauge theory in the commutative
%space, at least at weak coupling.

\section{Numerical results}

\subsection{Two dimensions}

\begin{figure}[htbp]
\vspace*{-4cm}
  \centering
  \includegraphics[width=1.1\linewidth]{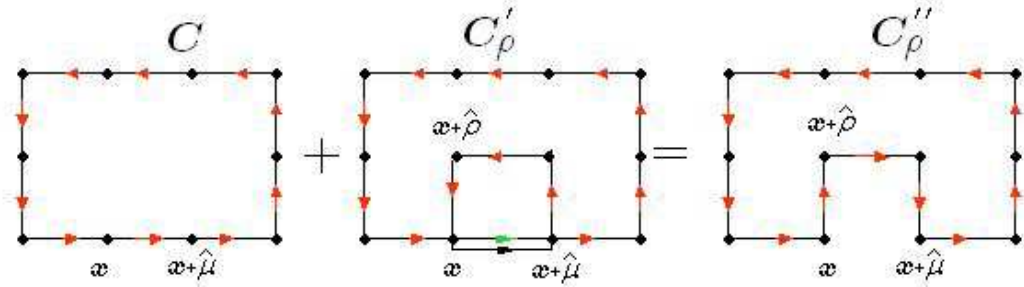}
\vspace*{-6cm}
\caption{\emph{An example for a set of contours, 
which are involved in a Schwinger-Dyson
equation. These equations relate the corresponding Wilson loops 
in the planar limit.}}
\label{SDcont}
%\vspace*{-3mm}
\end{figure}

\begin{figure}[htbp]
  \centering
  \includegraphics[width=.49\linewidth]{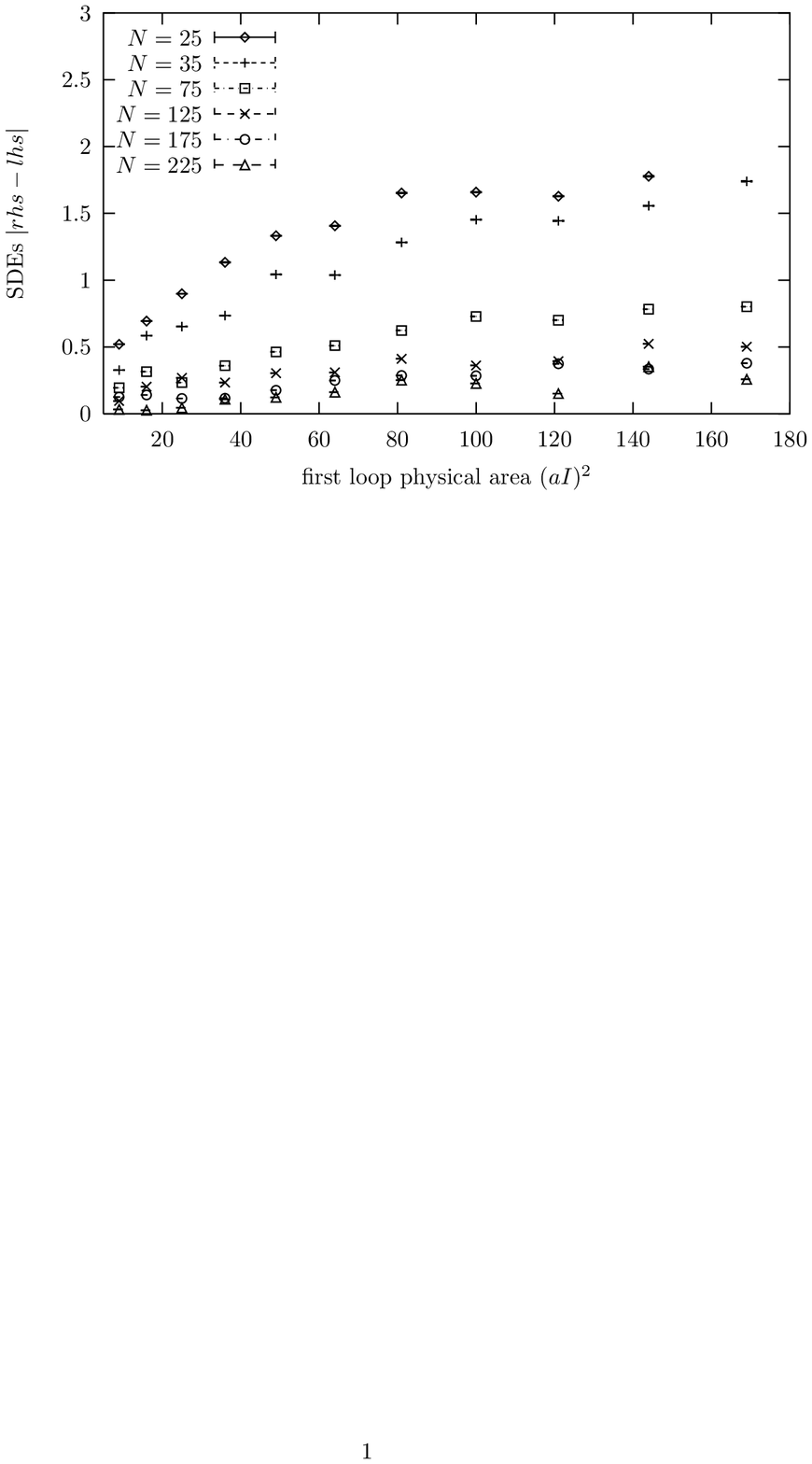} % \hspace*{4mm}
  \includegraphics[width=.49\linewidth]{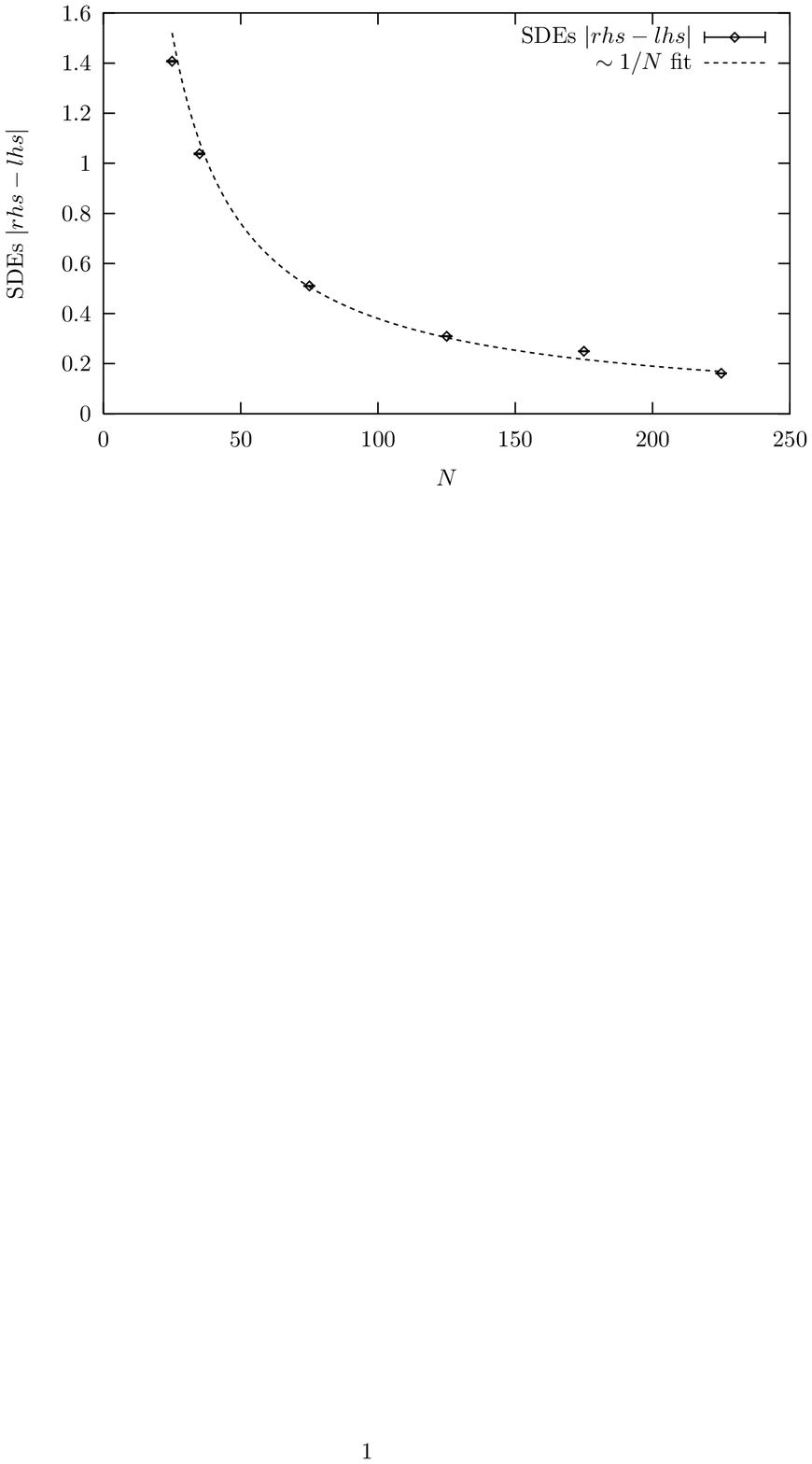}
  \caption{\emph{The convergence towards the validity of the Schwinger-Dyson 
equations as $N$ increases at fixed $\beta = 0.25$. On the $x$-axis 
of the plot on the left we show
the area of the contour called $C$ in Figure \ref{SDcont}. On the right we show
the convergence in $N$ at a fixed area of $7 \times 8$ for the contour $C$.
Note that in a common normalization --- with a prefactor $1/N$
for the Wilson loops --- our result is consistent
with the expected finite $N$ artifacts of the order of
$1/N^2$.}}
\label{SDfig}
\vspace*{-3mm}
\end{figure}

In the planar limit of the two dimensional model we obtain the
commutative $U( N \to \infty )$ lattice gauge theory on a plane,
which was solved by Gross and Witten %a long time ago 
by studying Wilson loops \cite{GroWit}. In this limit they found
an exact area law. We tested that the values of $N$, which are accessible
to our simulations, do approximate the planar limit well by checking
the validity of the large $N$ Schwinger-Dyson equations, along with
the large $N$ factorization. On the lattice, they
relate Wilson loops of different shapes in the planar limit.
An example for the corresponding contours is illustrated 
in Figure \ref{SDcont}. This was the property
that the original work by Eguchi and Kawai (without twist) was based on
\cite{EK}. Indeed, we observe that our measurements for the two sides of 
such equations converge as we increase $N$ to a magnitude of a 
few hundred, keeping $\beta $ fixed, for instance at $0.25$. 
This is shown in Figure \ref{SDfig}, see also Ref.\ \cite{NN}.

%\newpage
Hence we can use this limit to identify the lattice spacing
as $\beta \propto 1 /a ^{2}$, as in the result by Gross and Witten.
Therefore we take the double scaling
limit by keeping the ratio $N / \beta$ constant.

\begin{figure}[htbp]
  \centering
  \includegraphics[width=.48\linewidth]{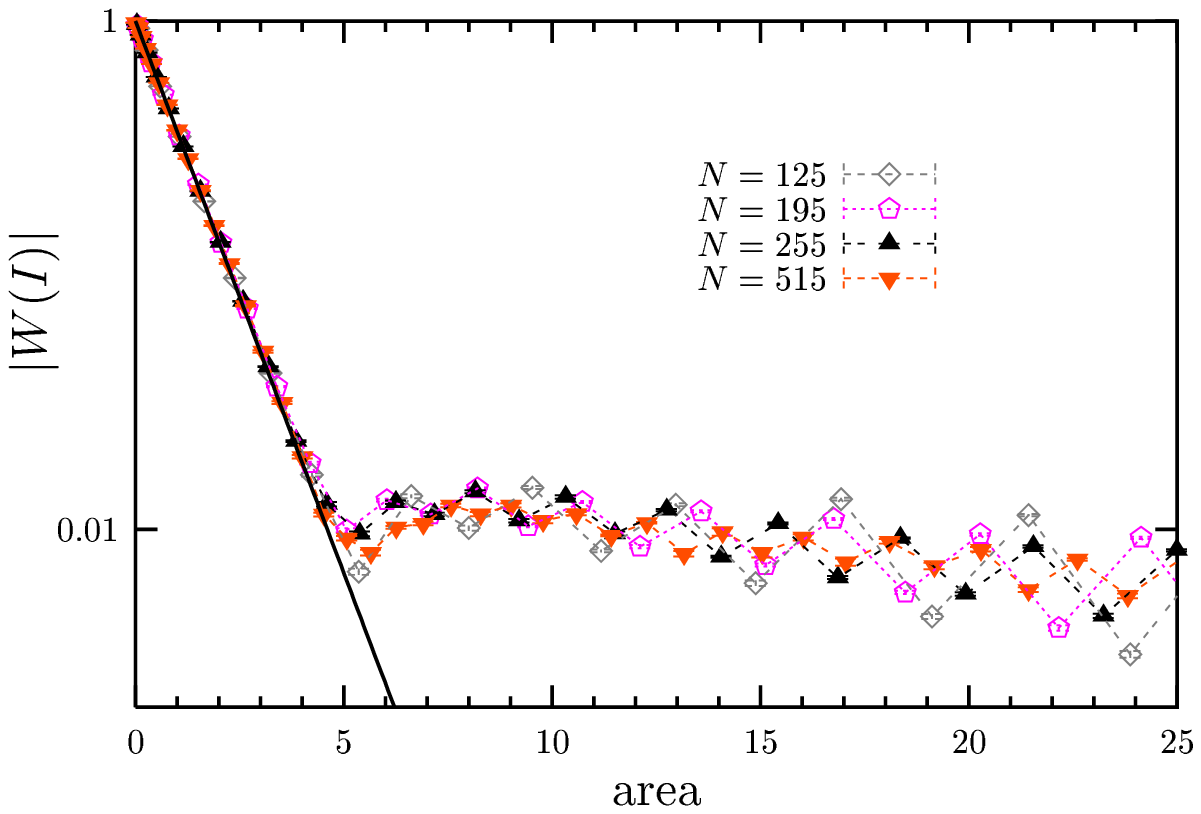} \hspace*{2mm}
  \includegraphics[width=.48\linewidth]{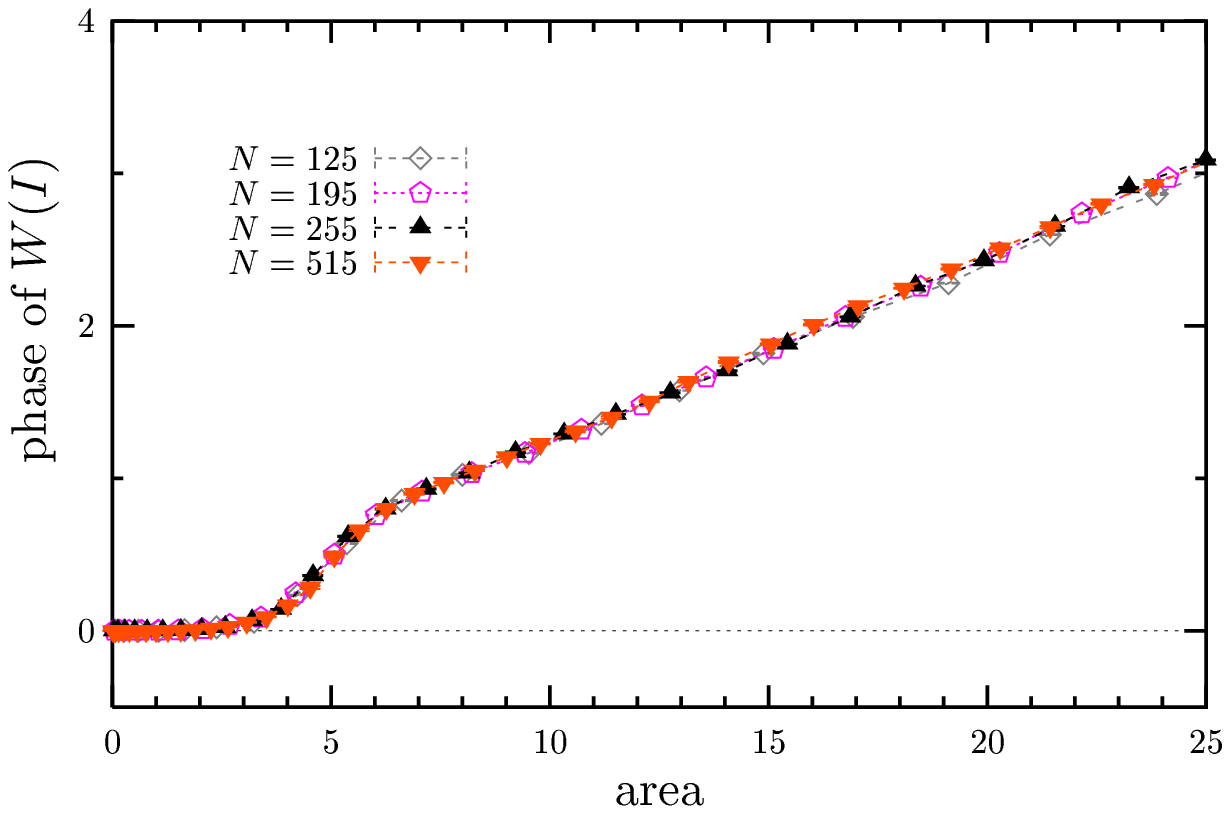}
\caption{\emph{We show square shaped Wilson loops $W(I)$
on a NC plane: from the plot
on the left we see that small loops follow the Gross-Witten area law
(which is marked by a line).
At large areas the absolute value does not decay any further, but a
linearly rising phase sets in, as the plot on the right shows.}}
\label{Wloop2d}
\end{figure}

In Figure \ref{Wloop2d} we show our simulation results \cite{NCQED2d}
for the Wilson loop 
\begin{equation}
W(I) = \langle W_{12}( I \times I) \rangle 
\end{equation}
in polar coordinates, as a function of the physical 
(i.e.\ dimensionful) area $(aI)^{2}$. We choose $N$ over a broad range,
with a fixed ratio $N / \beta = 32$.

From Figure \ref{Wloop2d} we infer the following properties:

\begin{itemize}

\item The observable $W(I)$ does indeed stabilize in the double scaling
limit. Thus we have found a new universality class. In particular
this shows that the model is non-perturbatively renormalizable.

\item At small area, the absolute value follows the Gross-Witten
area law, which is marked by a line in the plot on the left-hand-side 
of Figure \ref{Wloop2d}. In that regime the phase is practically zero,
hence Wilson loops with areas $A = (a I)^{2} \lsim 4$ follow the behavior
the behavior of commutative large $N$ gauge theory.

\item For larger areas, the absolute value does not decay any further, 
but the phase starts to increase linearly in the area.

\end{itemize}

\begin{figure}[htbp]
\vspace*{-7mm} %\hspace*{-3cm}
\subfigure{\epsfig{figure=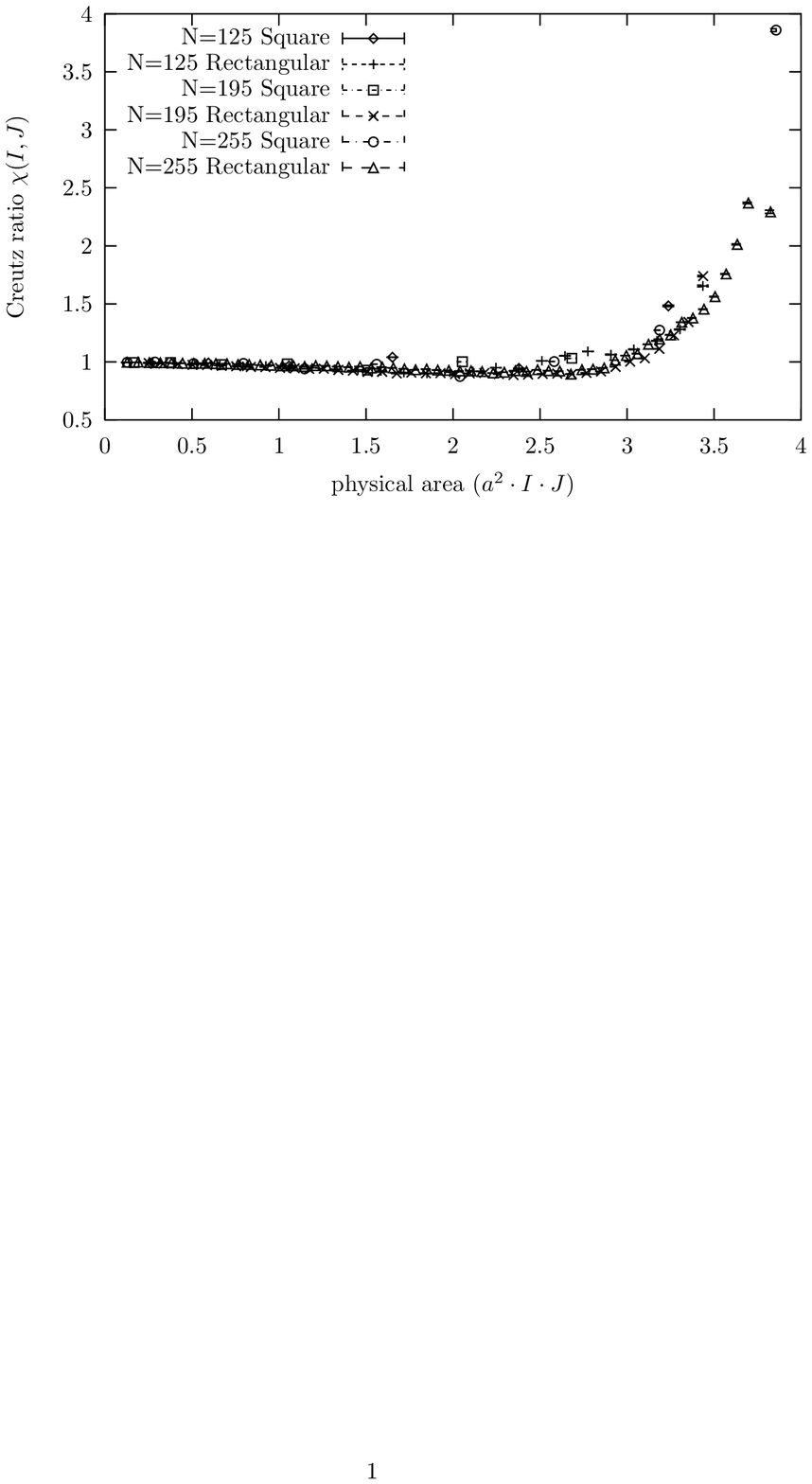,width=.55\linewidth}}%
%\vspace*{-10cm}
%\hspace*{-2cm}
\subfigure{\epsfig{figure=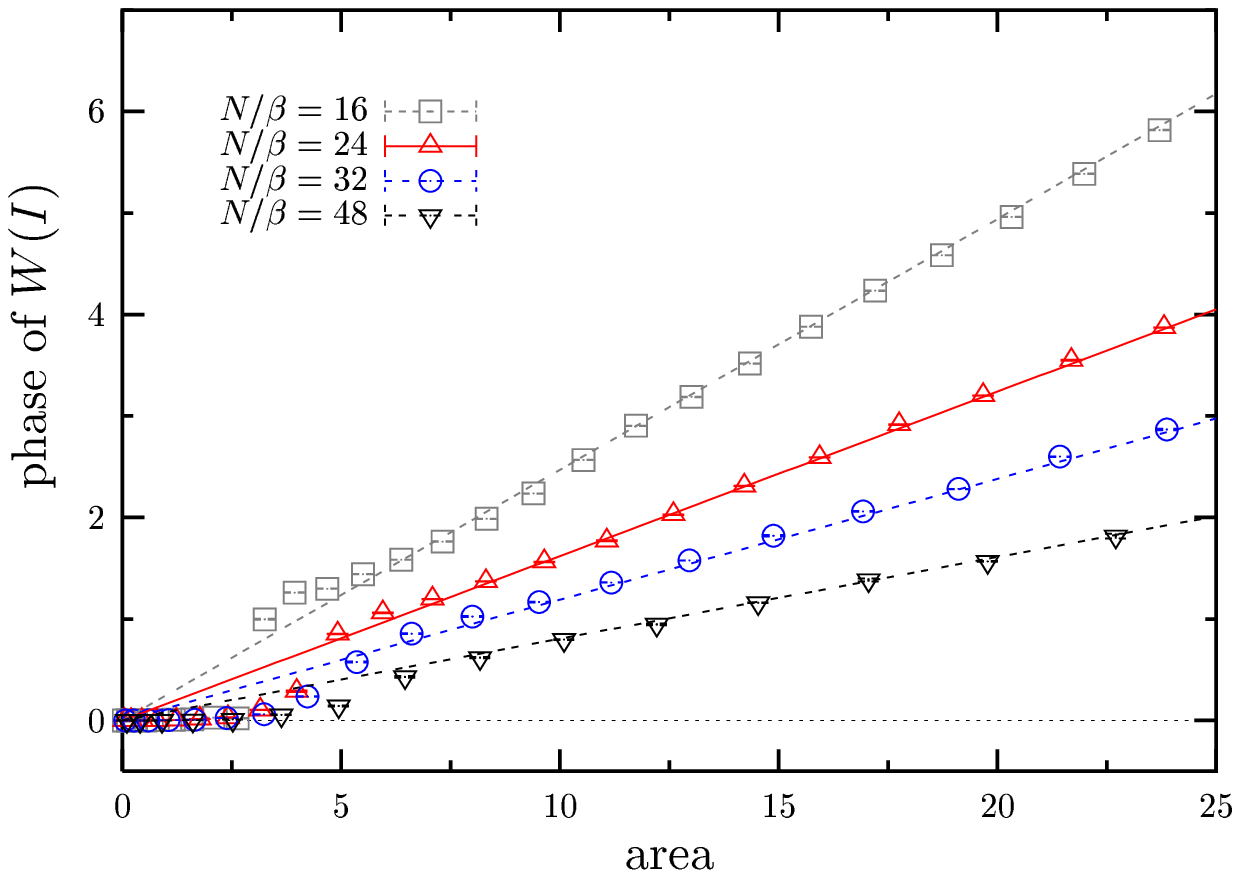,width=.45\linewidth}}
\vspace*{-5mm}
\caption{\emph{On the left: the Creutz ratio for Wilson loops with small areas,
up to the transition observed in Figure \ref{Wloop2d}. On the right: the 
linearly increasing phase at various values of the non-commutativity
parameter $\theta$, in agreement with eq.\ (\ref{ABeq}).}}
\label{Creutz_phasefig}
\end{figure}

Regarding the second point --- the area law for small Wilson loops ---
we also measured the Creutz ratio, 
\begin{equation}
\chi (I,J) = - \log \Big( \frac{W(I \times J) 
W((I-1) \times (J-1))}{W((I-1) \times J) W(I \times (J-1))} \Big) \ ,
\end{equation}
which singles out the string tension
$\sigma$ for decays $\propto \exp (- \sigma A)$.
Typical results for (nearly) square shaped Wilson loops, $\chi (I,I)$,
as well as extremely anisotropic (rectangular) Wilson loops, $\chi (2,J)$,
are shown in Figure \ref{Creutz_phasefig} on the left.
We see a stable value $\sigma \approx 1$ as long as we are well inside the
area law regime, and a marked deviation from it as the area approaches
its critical value.

In view of the third observation --- the linearly rising phase ---
we performed further tests with different ratios 
$N /\beta = 16, \ 24, \ 48$, i.e.\ different values of $\theta$,
see Figure \ref{Creutz_phasefig} on the right,
and also with rectangular loops.
At large areas we {\em always} found to a very high precision the 
simple relation
\begin{equation}  \label{ABeq}
{\rm Phase} = \frac{A}{\theta} = A \cdot B \ .
\end{equation}
In the last term we introduced a formal magnetic field $B = 1/\theta$,
in the spirit of Peierls' map.

This behavior corresponds exactly to the {\em Aharonov-Bohm effect}.
It is known from the corresponding magnetic fields in string
theory and in solid state physics, where it is implemented
on tree level. Here, however, we recover the very same behavior 
unexpectedly as a dynamical effect at low energy.

One may wonder why the short-ranged non-commutativity
has striking effects on the large rather than the small Wilson loops.
Apparently there is some UV/IR mixing \cite{UVIR} going on, even though
there are no divergences in the perturbative expansion of this model.
This suggests that UV/IR mixing occurs non-perturbatively, and it belongs
therefore to the fundamental nature of NC field theory.
This is in agreement with analytic \cite{phi4ana} and numerical 
\cite{phi4MC} results for the NC $\lambda \phi^{4}$ model.

\subsection{Four dimensions}

We now proceed to four dimensions, which include a NC plane, as well
as two commutative coordinates --- among them is the Euclidean time.
Therefore we deal with one NC plane, one commutative plane and 4 mixed planes,
and we will distinguish the (flat) Wilson loops accordingly.

There are 1-loop studies about the photon dispersion relation
in such spaces \cite{Mat00}. Computations of the $\beta$ function
indicate asymptotic freedom \cite{afree}.
Other perturbative studies suggest that this model may have a negative
IR divergence, which would render it unstable and unsuitable to describe
physics \cite{IRinstab}. More precisely, those works suggest that 
bosonic (fermionic)
degrees of freedom contribute negative (positive) IR poles, and they therefore
proceed to SUSY models. Here we stay in the simple bosonic framework; of course
we would like to verify this claim on a non-perturbative level.

Again we introduce a lattice formulation and map the NC plane on a twisted
Eguchi-Kawai model; one on each lattice point of the remaining (commutative) 
plane. On the technical side, it is very important to apply the heat bath
algorithm for the simulations. This is not straightforward, since this
matrix model is non-linear in the link variables.
For our 2d model a method to linearize it with the help of auxiliary fields
was introduced by Fabricius and Haan \cite{Fab84}. We also applied
it in our simulations, which led to the results in the previous Subsection.
In $d=4$ it turned out that this method is still applicable in a generalized
form \cite{privnote}, which allows us to use the powerful heat bath algorithm.
Without it, it is very hard to overcome the problems
related to thermalization and autocorrelation \cite{ProvVic}.

\begin{figure}[t]
\begin{center}
%\vspace{-30pt}
\scalebox{0.6}{\includegraphics{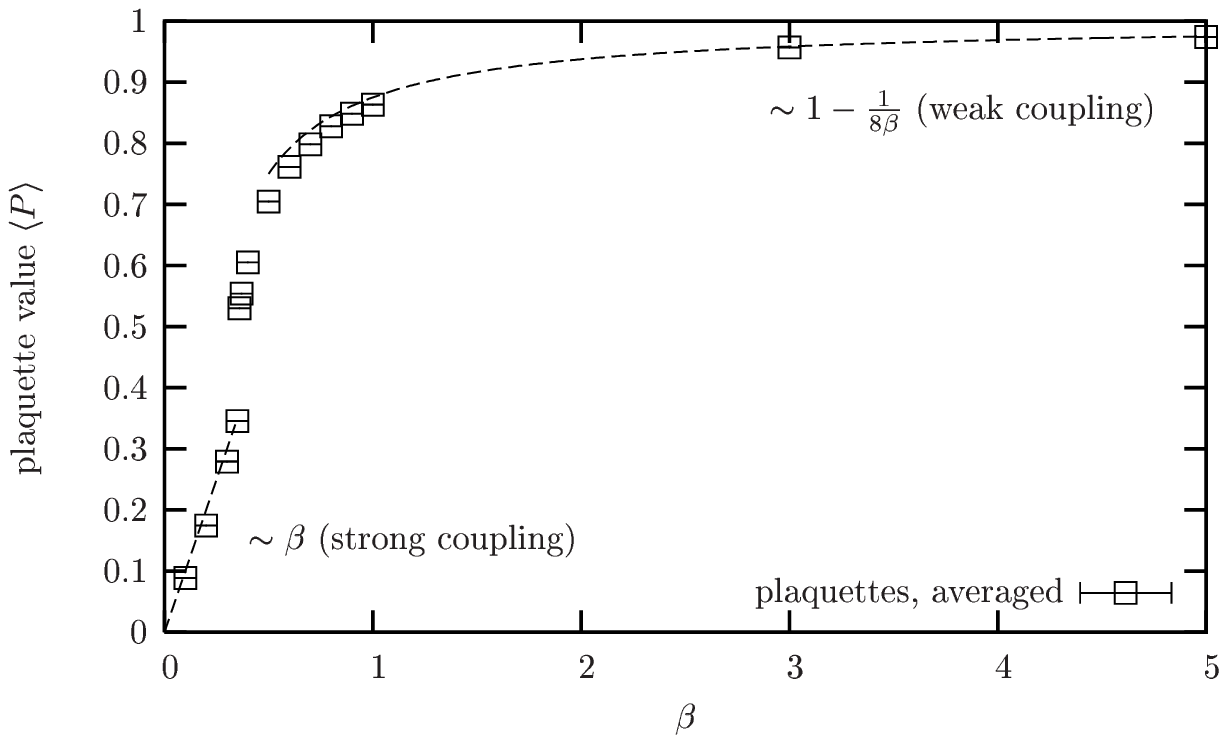}} 
\scalebox{0.6}
{\includegraphics{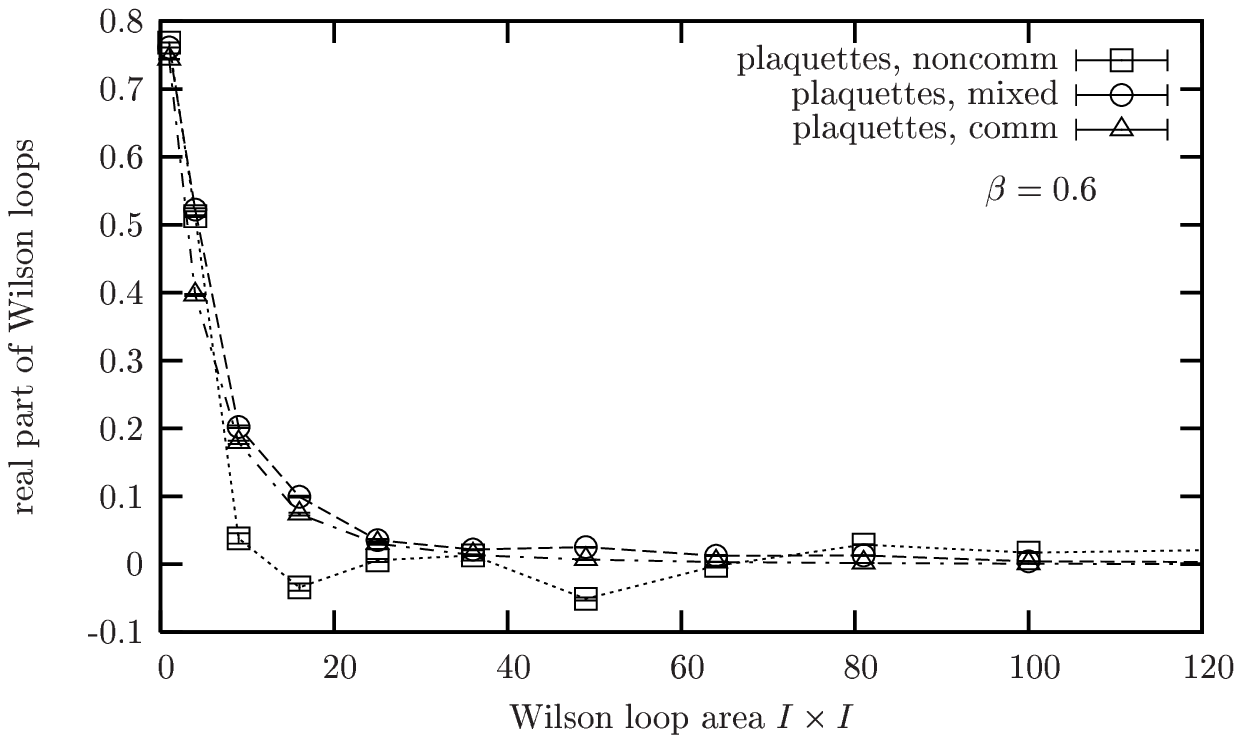}}
\vspace{-3mm}
\caption{\emph{On the left: 
The real part of the plaquettes, averaged over all possible 
orientations. The dashed lines show the asymptotic behavior expected
from expansion in strong resp.\ weak coupling.
On the right: The real part of the plaquettes, plotted separately 
according to their orientation.}}
\label{plaquettes}
\end{center}
%\vspace{-1cm}
\end{figure}

We first present simulation results at $N=25$ \cite{Lat04}.
\footnote{More precisely, this means that the lattice size is $25 \times 25$ 
in the NC plane (remember that this extent has to be odd) and $24 \times 24$
in the commutative plane (this is suitable for parallel computing).}
The left side of Figure \ref{plaquettes} shows the real part of the
plaquette values at different $\beta$. In this case all six possible orientations 
of the plaquettes are averaged over. The results match the 
anticipated asymptotic behavior \cite{TEK} of the plaquette expectation value
$\langle P \rangle$: at weak coupling is behaves as
$\sim 1-\frac{1}{8 \beta}$ and at strong coupling as $\sim \beta$.  
A phase transition occurs around $\beta \approx 0.35$.
All the results that we show in the following 
were obtained in the weak coupling phase.

The decay of square shaped Wilson loops with increasing area is shown in
Figure \ref{plaquettes} on the right. In that plot the results are 
separated according to the plaquette orientations (NC, mixed and commutative). 
The real part of the NC Wilson loops performs a damped oscillation around zero. 
This is in qualitative agreement with the observations in $d=2$. 
For the other two types of Wilson loops the real part
decays monotonously towards zero.

%\begin{figure}[t]
%\begin{center}
%\scalebox{0.6}{\includegraphics{creutz_avg.eps}}
%%\vspace{-40pt}
%\end{center}
%\caption{\emph{Creutz ratios $\chi$ of rectangular
%Wilson loops, averaged over all planes, at various couplings.}}
%\label{creutz}
%%\vspace{-1cm}
%\end{figure}

We have again measured the Creutz ratios for various rectangular Wilson loops
\cite{Lat04}. The (dimensionless) string tension approaches zero as 
$\beta \rightarrow \infty$.

\begin{figure}[t]
\begin{center}
\scalebox{0.6}{\includegraphics{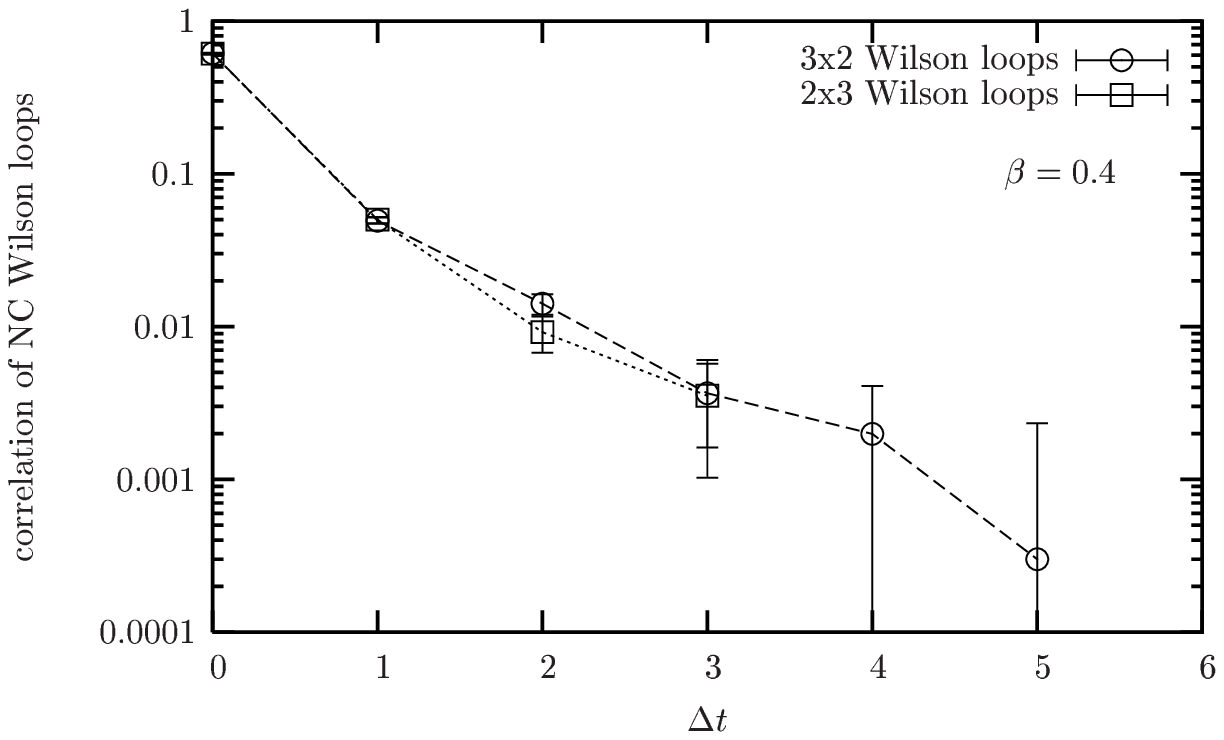}}
\scalebox{0.6}{\includegraphics{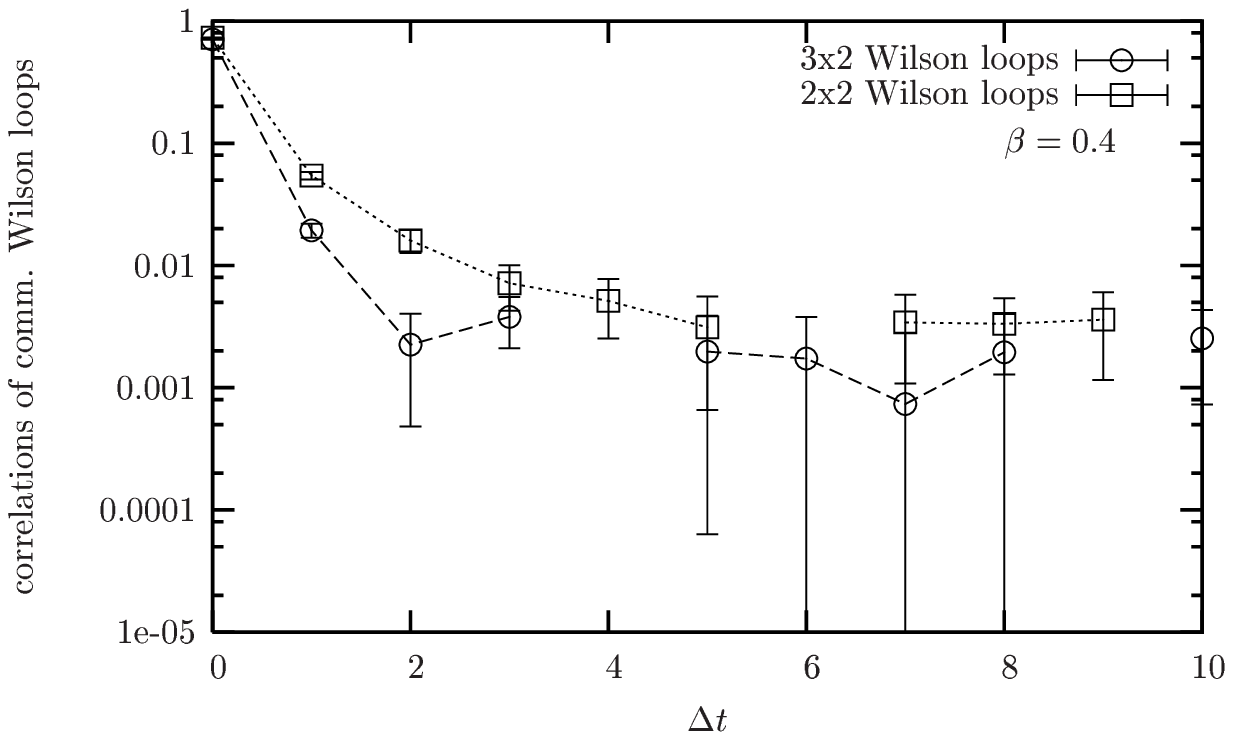}}
%\vspace{-40pt}
\end{center}
\vspace*{-5mm}
\caption{\emph{Temporal correlation of the plaquettes lying completely 
in the NC plane (on the left) and completely in the commutative plane
(on the right). In the former case only we see a trend towards 
an exponential decay.}}
\vspace{-2mm}
\label{corre}
\end{figure}

Figure \ref{corre} (on the left) shows the correlation function of the 
plaquettes lying completely in the NC plane, separated by $\Delta t$ 
in Euclidean time. These data were taken at 
$\beta=0.4$, i.e.\ barely in the weak coupling phase. 
The decay is likely to be exponential, but a higher statistics is 
required to verify this behavior.

The analogous plot for plaquettes in the commutative plane is shown in 
Figure \ref{corre} on the right. In that case the decay does not 
seem exponential.
 
\begin{figure}[hbt]
\begin{center}
\hspace*{1cm}
\def\fpsangle{270}
\epsfxsize=50mm
\fpsbox{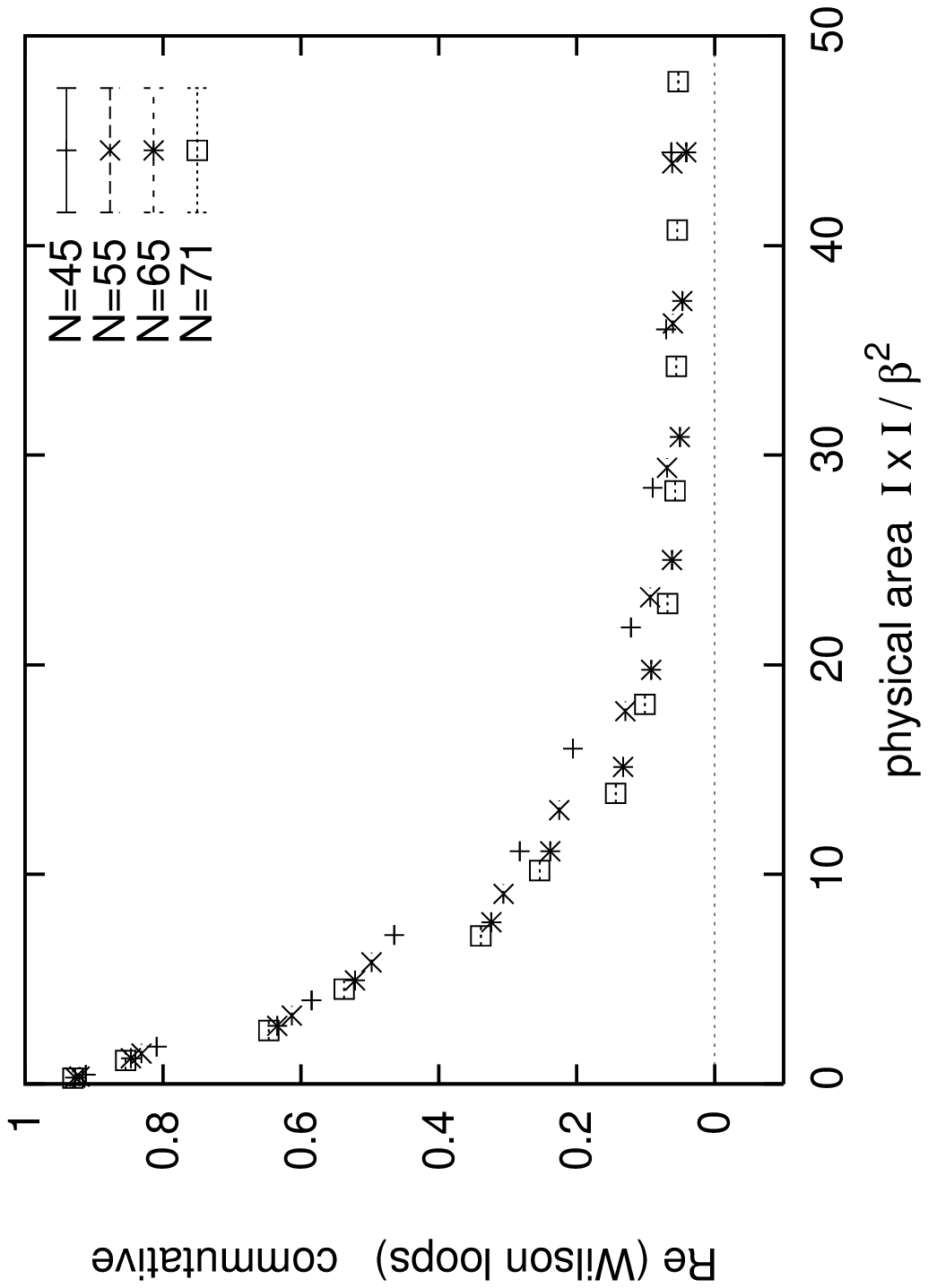}
%\vspace*{-1cm}
\def\fpsangle{270}
\epsfxsize=50mm
%\hspace*{1cm}
\fpsbox{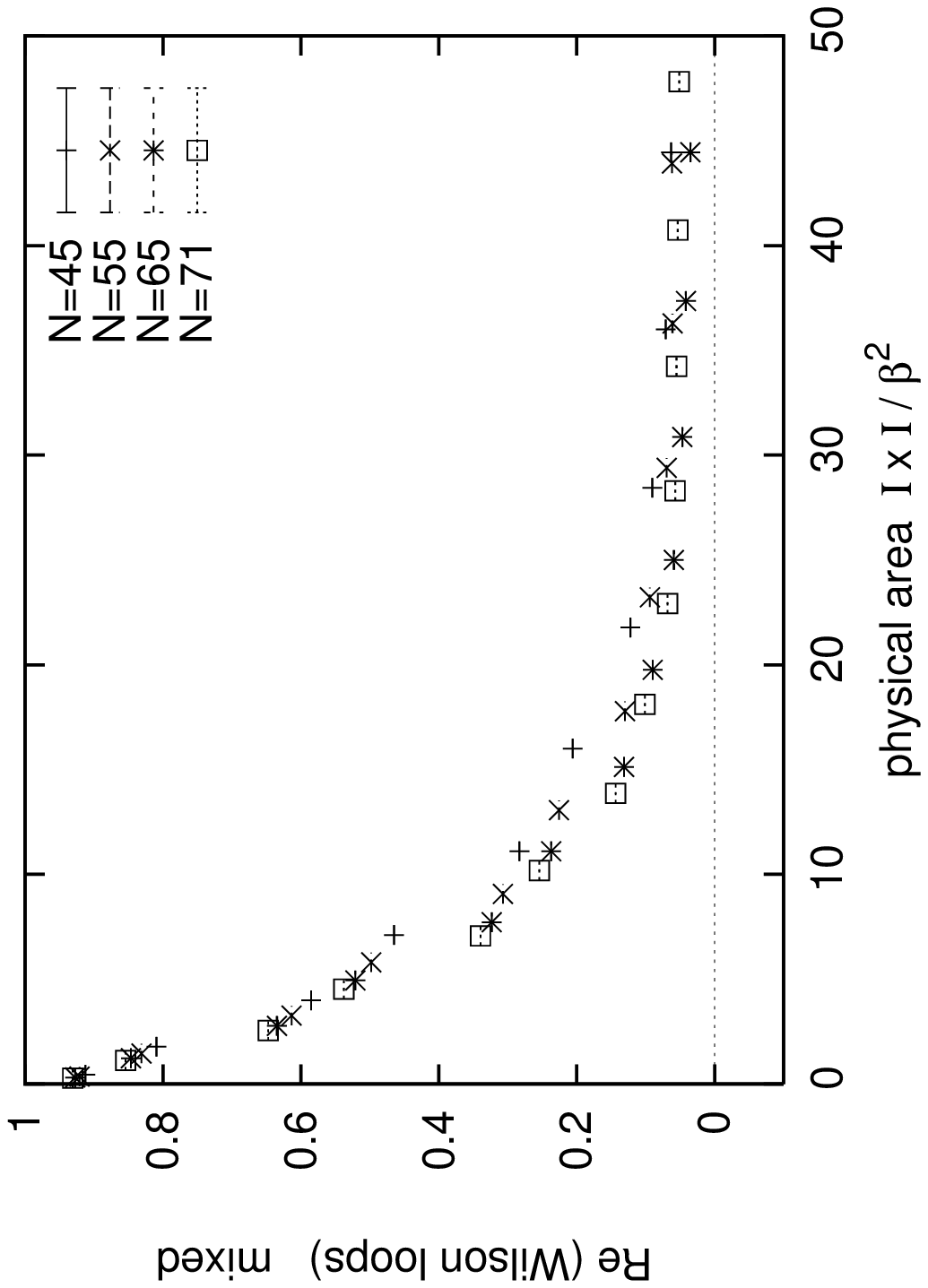}
\def\fpsangle{270}
\epsfxsize=50mm
%\hspace*{1cm}
\fpsbox{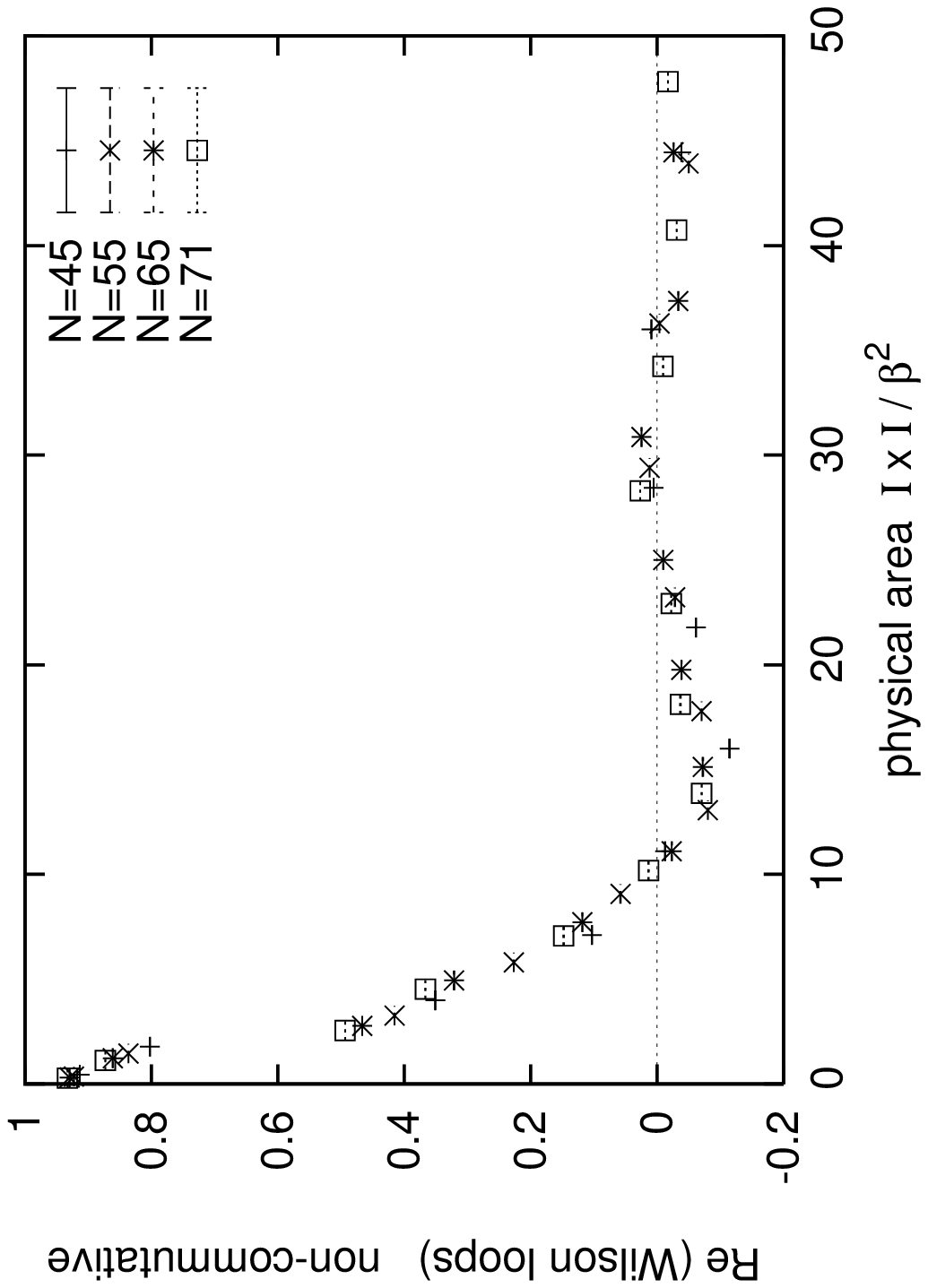}
\end{center}
\vspace{-3mm}
\caption{\emph{We show the real part of Wilson loops in the three types of planes
(commutative, mixed and NC) of our 4d space. Here we have fixed
$N / \beta^2 =20$. The results approximate a double scaling behavior 
decently well. (We always refer to $N\times N$ matrices for the NC plane,
and a $(N-1)\times (N-1)$ lattice on the commutative plane.)}}
\label{DSL4d}
\vspace*{-4mm}
\end{figure}

At last, we illustrate our progress in the search for a double scaling limit.
In contrast to $d=2$, we do not have an analytic result in the planar
limit to identify the physical lattice spacing (i.e.\ to relate $\beta$ and $a$).
However, we observed that double scaling is approximated quite 
well if we assume ad hoc the simple
relation $\beta \propto 1/a$, so that the double scaling limit keeps
$N / \beta^{2} = const.$ In particular we set this ratio to $20$
and obtained the results for the real parts of square shaped Wilson loops
shown in Figure \ref{DSL4d}. We are also working on a systematic search
for double scaling by matching the data without any assumption about
the relation between $\beta$ and $a$. This represents a completely 
unbiased test of the result presented here.

\section{Conclusions}

We have simulated NC $U(1)$ field theory in $d=2$ and in $d=4$.
In $d=2$ we observed an area law for Wilson loops with small
area, as expected from the equivalence to commutative large $N$
gauge theory in the UV regime. At large area, a phase
starts to increase linearly in the area, which is reminiscent of
the Aharonov Bohm effect.
 
The simulations in $d=4$ (with two NC coordinates) are still on-going.
We first measured the action, plaquettes, Wilson loops and Creutz 
ratios. The results for the action and plaquettes agree 
with asymptotic predictions. There seems to be a vanishing string 
tension at weak coupling in the part of the parameter space we explored. 
The Wilson loops in the NC plane show a behavior similar to the
2d NC model, whereas the loops in other planes are close to the
commutative feature.

We found a simple ad hoc ansatz which seems to provide quite well
a double scaling limit. This ansatz is still being checked with
a fully systematic approach. The double scaling limit corresponds 
to the model at zero lattice spacing and in infinite volume,
hence its identification is a milestone in the non-perturbative 
analysis of this model.
Thus we hope to glimpse at the dispersion relations for the NC photon, 
which might allow us to confront the theory with phenomenological data.

Even before that, however, it is important to clarify if this model
is actually IR stable.

\vspace*{3mm}
%%%%%%%%%%%%%%%%%%%%%%%%%%%%%%%%%%%%%%%%%%% 
%{\small
{\bf Acknowledgement} It is a pleasure to thank Jan Ambj\o rn, 
Aiyalam Balachandran, Simon Catterall,
Harald Dorn, Giorgio Immirzi,
Esperanza Lopez, Yuri Makeenko, Denjoe O'Connor, Richard Szabo 
and Alessandro Torrielli for stimulating discussions.
We are also grateful to the ``Deutsche Forschungsgemeinschaft'' (DFG) for 
generous support. An important 
part of the computations were performed on the IBM p690 
clusters of the ``Norddeutscher Verbund f\"ur Hoch- und 
H\"ochstleistungsrechnen'' (HLRN). Finally we thank Hinnerk St\"{u}ben for 
his advice on the parallelization of our codes.
%}
%%%%%%%%%%%%%%%%%%%%%%

\end{document}